\documentclass[final,5p,times]{elsarticle}
 \usepackage{color,graphicx,epsfig}
 \usepackage{amsmath,amssymb,dsfont,epsfig,graphicx,xcolor}
 \usepackage{ifpdf}
 \usepackage{amsmath}
 \usepackage{bm}
 \usepackage{color}
 \usepackage[english]{babel}
 \usepackage{graphicx}%
 \usepackage{amsfonts}%
 \usepackage{amssymb}
 \usepackage{braket}
 \usepackage{hyperref}
 \usepackage{epstopdf}
 \usepackage{cancel}
 \usepackage[capitalise]{cleveref}
 \usepackage{booktabs}
 \usepackage{diagbox}
 \usepackage{multirow}
 \graphicspath{{Plots/}}

\definecolor{nicered}{rgb}{0.7,0.1,0.1}
\definecolor{nicegreen}{rgb}{0.1,0.5,0.1}
\definecolor{violet}{rgb}{0.7,0.3,0.3}
\hypersetup{colorlinks,citecolor= nicegreen,linkcolor= nicered}

\newcommand{\nc}{\newcommand}
\nc{\non}{\nonumber}{}
\nc{\hc}{\hbox {H.c.}}
\nc{\noi}{\noindent}
\nc{\barx}{\bar{x}}
\nc{\pbarn}{\;\hbox {pb}}
\nc{\fbarn}{\;\hbox {fb}}

\nc{\hsp}{\hspace{0.5cm}}
\nc{\lsp}{\hspace{1cm}}
\nc{\Lsp}{\hspace{2cm}}
\nc{\LLsp}{\lsp\lsp}
\nc{\lra}{\longrightarrow}
\nc{\p}{\prime}
\nc{\sgn}{\text{sgn}}
\nc{\ph}{\varphi}
\nc{\op}{{\cal O}}
\nc{\eq}{\text{Eq.~}}

\newcommand{\abs}[1]{\left\lvert#1\right\rvert}

\nc{\beq}{\begin{equation}}  \nc{\eeq}{\end{equation}}
\nc{\bea}{\begin{eqnarray}}  \nc{\eea}{\end{eqnarray}}
\nc{\baa}{\begin{array}}     \nc{\eaa}{\end{array}}
\nc{\bit}{\begin{itemize}}   \nc{\eit}{\end{itemize}}
\nc{\ben}{\begin{enumerate}} \nc{\een}{\end{enumerate}}
\nc{\bce}{\begin{center}}    \nc{\ece}{\end{center}}
\nc{\bpm}{\begin{pmatrix}}   \nc{\epm}{\end{pmatrix}}
\nc{\bvt}{\begin{verbatim}}  \nc{\evt}{\end{verbatim}}

\def\LjubljanaFMF{Faculty of Mathematics and Physics, University of Ljubljana,
 Jadranska 19, 1000 Ljubljana, Slovenia }
\def\LjubljanaIJS{Jo\v zef Stefan Institute, Jamova 39, 1000 Ljubljana, Slovenia}
\def\Orsay{Universit\'e Paris-Saclay, CNRS/IN2P3, IJCLab, 91405 Orsay, France}
\def\Bari{Istituto Nazionale di Fisica Nucleare, Sezione di Bari, Via Orabona 4, I-70126 Bari, Italy}

\makeatletter
\def\ps@pprintTitle{%
  \let\@oddhead\@empty
  \let\@evenhead\@empty
  \def\@oddfoot{\reset@font\hfil\thepage\hfil}
  \let\@evenfoot\@oddfoot
}
\makeatother

\biboptions{sort&compress}

\begin{document}

\begin{frontmatter}
\title{Probing CP violation in exclusive $b \to s \nu \bar \nu$ transitions}

\author[Orsay]{S\'ebastien Descotes-Genon}
\ead{sebastien.descotes-genon@ijclab.in2p3.fr}

\author[LjubljanaIJS,LjubljanaFMF]{Svjetlana Fajfer}
\ead{svjetlana.fajfer@ijs.si}

\author[LjubljanaIJS,LjubljanaFMF]{Jernej~F.~Kamenik}
\ead{jernej.kamenik@cern.ch}

\author[Bari]{Mart\'in Novoa-Brunet\corref{mycorrespondingauthor}}
\cortext[mycorrespondingauthor]{Corresponding author}
\ead{martin.novoa@ba.infn.it}

\address[Orsay]{\Orsay}
\address[Bari]{\Bari}
\address[LjubljanaIJS]{\LjubljanaIJS}
\address[LjubljanaFMF]{\LjubljanaFMF}

\begin{abstract}
We consider the time-dependent analysis of rare $B_d$ and $B_s$ decays mediated by $b\to s \nu \bar \nu$ transitions. The inclusion of time evolution allows us to construct novel observables with specific sensitivity to CP-odd phases in these processes. The sensitivity to CP violation of corresponding time-integrated measurements in presence of flavor-tagging is also explored. We provide precise predictions for these observables in the SM and explore their sensitivity to new CP-violating NP contributions at present and planned future $B$-physics experiments. As such, these observables provide unique probes of CP violation in $b \to s \nu \bar \nu$ transitions.
\\[0.4cm]Report Number: BARI-TH/22-735
\end{abstract}

\end{frontmatter}

\section{Introduction}\label{sec:introduction}

Recent experimental data in $B$ physics hint toward deviations from Lepton Flavour Universality (LFU) in semi-leptonic decays~\cite{Bifani:2018zmi} at significances from 2.3$\sigma$ to 2.6$\sigma$ as measured by LHCb~\cite{Aaij:2014ora,Aaij:2019wad,Aaij:2017vbb}. Belle has also recently reported measurements of $R_K$~\cite{Abdesselam:2019lab} and $R_{K^*}$~\cite{Abdesselam:2019wac} in agreement with LHCb measurements, but with much larger uncertainties.
In addition to these LFU ratios, LHCb data exhibit deviations close to $3\sigma$ from the Standard Model (SM) expectation in the $P^\prime_5$ angular observable of $B \to K^*\mu\mu$ decay~\cite{Descotes-Genon:2013vna}, and milder deviations are also seen in branching ratios of $b\to s\mu\mu$ exclusive decays~\cite{Aaij:2013qta,Aaij:2014pli,Aaij:2015oid,Aaij:2020nrf,Aaij:2013aln}. Deviations are also hinted at in Belle data for $B\to K^*\mu\mu$~\cite{Abdesselam:2016llu,Wehle:2016yoi}. These deviations can be explained in a very economical way through New Physics (NP) contributions to Wilson coefficients for vector/axial operators describing $b\to s\ell^+\ell^-$ at the scale $\mu=m_b$, as shown by global fits to $b\to s\gamma$, $b\to see$ and $b\to s\mu\mu$ observables (see e.g Refs.~\cite{Alguero:2021anc,Altmannshofer:2021qrr}).

An important complementary probe of NP is provided by $b\to s\nu \bar \nu$ transitions, since the SM neutrinos reside in the same leptonic weak doublets as the left-handed charged leptons. In particular, the decays $B\to h_s \nu \bar \nu$, with $h_s$  standing  for  hadronic states of unit strangeness, are known for their NP sensitivity~\cite{Altmannshofer:2009ma}.  In the SM,  branching ratios are found to be $\mathcal B(B\to K^{(*)}  \nu \bar \nu)_\mathrm{SM}=(9.6 \pm 0.9)\times 10^{-6}$ and $\mathcal B(B^+ \to K^+  \nu \bar \nu)_\mathrm{SM}= (5.6\pm 0.5)\times 10^{-6}$ \cite{Kou:2018nap} (a recent update can be found in Ref.~\cite{Buras:2022wpw}). The Belle collaboration has produced limits at 90\% Confidence Level (CL): $\mathcal B(B^0\to K^{*0}\nu \bar \nu)_\mathrm{exp}< 1.8 \times 10^{-5}$,   $\mathcal B(B^+\to K^{*+}\nu \bar \nu)_\mathrm{exp}< 6.1 \times 10^{-5}$ and  $\mathcal B(B^+ \to K^+  \nu \bar \nu)_\mathrm{exp}< 1.9\times 10^{-5}$ \cite{Grygier:2017tzo}. Recently, the Belle II collaboration also presented a first bound of $\mathcal B(B^+\to K^{*+}\nu \bar \nu)_\mathrm{exp}< 4.1 \times 10^{-5}$~\cite{Dattola:2021cmw} using 63\,fb${}^{-1}$ of collected luminosity. It plans to observe these three decay modes  with about $10$ ab$^{-1}$, while the sensitivities to the SM branching ratio will reach a precision of about $10\%$ with $50$ ab$^{-1}$~\cite{Kou:2018nap}. Assuming a signal efficiency of the same order $\mathcal O(10^{-3})$ at Belle II as in Belle, one thus expects $\mathcal O(300)$ events from $B^+ \to K^+ \nu\bar\nu$ and $B^0 \to K^{*0}\nu\bar\nu$ in the SM at $50$\,ab${^{-1}}$ with a signal-over-background ratio $S/B \sim 0.3$. Some preliminary sensitivity studies are also available for proposed future high energy $e^+ e^-$ colliders, such as the CEPC~\cite{CEPCStudyGroup:2018ghi} and FCC-$ee$~\cite{FCC:2018evy}. In particular, a recent study of the $B_s \to \phi \nu\bar\nu$ mode at CEPC~\cite{Li:2022tov} is projecting a signal efficiency of almost $2.5\%$ resulting in around $7500$ signal events per $10^{12}$~Z decays and $S/B \sim 1$. Scaling these numbers to the planned FCC-$ee$ luminosities at the $Z$-pole~\cite{FCC:2018byv} one can project up to $\mathcal O(20k)$ reconstructed $B_s \to \phi \nu\bar\nu$ decays. Such statistics could allow for precision studies of rare $b\to s \nu\bar \nu$ transitions well beyond the measurement of branching ratios.

Until recently, the global fits to $b\to s\ell^+\ell^-$ data have mostly focused on NP scenarios with purely real contributions to the relevant Wilson coefficients~\cite{Alguero:2021anc}. The assumption is reasonable if NP closely follows the same pattern of flavour (and CP) breaking as the SM, i.e. adheres to Minimal Flavour Violation (MFV) in the quark sector~\cite{DAmbrosio:2002vsn}. However, beyond MFV, there is no guarantee for NP contributions to be aligned and thus purely real (relative to SM), and one could imagine sizeable NP-induced CP-violating effects in these decays. Due to the large uncertainties on currently measured CP-odd observables, such contributions are only loosely constrained, as illustrated by recent global fits considering  complex NP scenarios~\cite{Altmannshofer:2021qrr}, suggesting the need for better measurements and additional observables to probe CP-violation in these transitions. Let us highlight Ref.~\cite{Becirevic:2020ssj}, where the measurement of the CP asymmetries below and above the peaks of the $J/\psi$ and $\psi(2s)$ resonances was suggested in order to reveal the presence of NP-induced complex Wilson coefficients in $B \to K \mu^{+} \mu^{-}$.

These considerations lead to the question whether such CP-violating contributions could also arise and be probed in $b\to s\nu\bar\nu$ transitions. As we show in the following, the observables discussed so far in the literature probe the corresponding phases only in the presence of right-handed currents, whose existence and size is yet to be confirmed. We thus propose an additional set of observables applicable to $B_d$ and $B_s$ decays, which are able to probe NP phases even in the absence of right-handed currents. These observables involve the interference between neutral-meson mixing and $b\to s\nu \bar \nu$ decay amplitudes, arising both at the loop level in the SM, and
normalized CP-asymmetries can be constructed with significantly reduced  hadronic uncertainties. 

Such observables have been recently discussed in the context of $b\to s\ell^+ \ell^-$ decays in Refs.~\cite{Descotes-Genon:2015hea,Descotes-Genon:2020tnz}, and we adapt a similar approach for $b\to s\nu \bar \nu$ transitions.
 In the case of coherent $B - \bar B$ production (at $B$-factories), these observables can be reconstructed through a time-dependent analysis, whereas the incoherent production can be exploited through time-integrated observables in presence of initial $B$ flavor tagging. The latter approach proves highly challenging for the current experimental environments such as LHC but might be feasible at future $Z$-factories like CEPC and FCC-$ee$. 

The remainder of the article is structured as follows: In Sec.~\ref{sec:btosnunu}, we recall the basics regarding the treatment of $b\to s\nu \bar \nu$ transitions in the weak effective Hamiltonian and reproduce the expressions for usual observables such as branching ratios and final state hadron polarisation fractions. In Sec.~\ref{sec:probingNPweakphases}, we introduce new observables probing the interference between mixing and decay for the case of $B_d\to K_S\nu\bar\nu$, $B_d\to K^{*0}(\to K_S\pi^0)\nu\bar\nu$ and $B_s\to \phi\nu\bar\nu$. In Sec.~\ref{sec:numerical}, we discuss the prospects for the measurement of these observables at present and planned $B$ physics experiments and what they can teach us about the presence of CP-odd NP phases in $b\to s\nu\bar\nu$, before concluding in Sec.~\ref{sec:conclusions}.

\section{Weak effective Hamiltonian and standard  $b\to s\nu\bar\nu$ observables}\label{sec:btosnunu}

The effective Hamiltonian relevant for $b \to s \nu \bar \nu$ transitions at the scale $\mu=m_b$  reads~\cite{Altmannshofer:2009ma}
\begin{equation}\label{eq:hameffbsnunu}
{\cal H}_{\rm eff}=-\frac{4 G_F}{\sqrt{2}} V_{tb}V_{ts}^* \sum_\nu (C_{L}^\nu {\cal O}_{L}^\nu + C_{R}^\nu {\cal O}_{R}^\nu) + {\rm h.c.} \,,
\end{equation}
with 
\begin{equation}
 {\cal O}_{L,R}^\nu = \frac{e^2}{16\pi^2} (\bar{s}\gamma_\mu P_{L,R} b)(\bar\nu\gamma^\mu(1-\gamma_5)\nu) \,,
\end{equation}
and $P_{L,R} =(1\mp\gamma_5)/2$. We assume that NP contributes significantly only through vector/axial operators (such as in NP scenarios currently favoured by global fits to $b\to s \ell^+\ell^-$ data) that do not entail CP-even (``strong'') phases, and that (anti)neutrinos produced in these decays are purely (right) left-handed. We have $C_{L,R}^{\nu} = C^{\nu,{\rm SM}}_{L,R} + C^{\nu,{\rm NP}}_{L,R} $ with $C^{\nu,{\rm SM}}_L = -6.38$ and $C^{\nu,{\rm SM}}_R=0$, with the same value for all three neutrino flavours.

Usual observables for the rare decays $B \to K^{(*)} \nu\bar\nu$ can then be conveniently expressed as~\cite{Altmannshofer:2009ma,Buras:2014fpa}
\begin{align}
\mathcal B(B \to K \nu \bar \nu ) =&  \mathcal B(B \to K \nu \bar \nu )_{\rm SM} \times \frac{1}{3}\sum_\nu (1 - 2 \eta_\nu) \epsilon_\nu^2\,,  \nonumber \\
\mathcal B(B \to K^* \nu \bar \nu ) =& \mathcal B(B \to K^* \nu \bar \nu )_{\rm SM} \times \frac{1}{3}\sum_\nu (1 + 1.31 \eta_\nu) \epsilon_\nu^2\,,  \nonumber \\
\mathcal B(B \to X_s \nu \bar \nu ) =&  \mathcal B(B \to X_s \nu \bar \nu )_{\rm SM}  \times \frac{1}{3}\sum_\nu(1 + 0.09 \eta_\nu) \epsilon_\nu^2\,,  \nonumber \\
\langle F_L \rangle =&  \langle F_L \rangle_{\rm SM} \times \frac{\sum_\nu (1 + 2 \eta_\nu)\epsilon_\nu^2}{\sum_\nu (1+1.31 \eta_\nu)\epsilon_\nu^2}\,,
\label{eq:BKnunu}
\end{align}
where $\langle F_L \rangle$ is the longitudinal $K^*$ polarisation fraction in $B\to K^* \nu \bar \nu$ decays. Updated predictions within the SM can be found in Ref.~\cite{Buras:2022wpw}.
For each flavour of neutrino $\nu=\nu_e,\nu_\mu,\nu_\tau$, the two NP parameters can in turn be expressed as
\beq
\epsilon_\nu = \frac{\sqrt{|C^\nu_L|^2 + |C^\nu_R|^2}}{|C^\nu_{\rm SM}|}\,, ~ \eta_\nu = \frac{- {\rm Re}(C^\nu_L C^{\nu*}_R)}{|C^\nu_L|^2 + |C^\nu_R|^2}\,,
\eeq
Note that any deviations from SM in $\langle F_L \rangle$ or non-universal deviations in $\mathcal B(B \to (K,K^*,X_s) \nu \bar \nu ) / \mathcal B(B \to (K,K^*,X_s) \nu \bar \nu )_{\rm SM}$ would signal the presence of right-handed quark currents ($\eta_\nu\neq0$). Conversely, in absence of right-handed currents, the above observables yield only $\epsilon_\nu$, which depends on the moduli of the Wilson coefficients $C_L^\nu$, but not their phases.

\section{Probing CP-odd phases through  interference between mixing and decay}\label{sec:probingNPweakphases}


We can thus conclude that these $b\to s\nu\bar\nu$ observables probe NP phases only through $\eta_\nu$, and only in the presence of right-handed currents (i.e. $C_R^\nu$). However, the existence of significant right-handed currents is currently far from certain: for instance, most NP scenarios favoured by $b\to s\ell^+\ell^-$ data do not feature them. One can consider another usual probe of CP-violation, i.e., direct CP-asymmetries comparing the rates of CP-conjugated modes (such as ${\mathcal B}(B^+\to K^+\nu\bar\nu)$ and ${\mathcal B}(B^-\to K^-\nu\bar\nu)$). However these asymmetries are expected to vanish in the absence of mechanisme able to generate significant strong phases. 

However, we can use a third handle on CP-violating New Physics, namely  the interference between neutral meson mixing and decay, though the time dependence of the decays $B_d \to K_S \nu\bar\nu$, $B_s\to \phi \nu\bar\nu$ or $B_d\to K^{*0}(\to K_S\pi^0)\nu\bar\nu$. The resulting phenomenology is analogous to standard measurements of (indirect) CP-violation routinely performed  at $e^+ e^-$ $B$-factories and hadronic machines (i.e. LHC) for example to measure the CKM unitarity triangle angles $\beta_{(s)}$. 

Let us emphasize that the time dependence studied is related to the evolution of the neutral mesons through mixing, but that its exact meaning depends on the experimental setup. In the case of coherent production of two neutral mesons through an $\Upsilon$ resonance decay (at $e^+ e^-$ $B$ factories like Belle II), the measurements are typically performed as a function of $\Delta t$ which is the time difference between the decays of the two neutral mesons (ranging from $-\infty$ to $+\infty$, with an exponential factor of the form $\exp(-\Gamma |t|)$). On the other hand, in the case of an incoherent production and subsequent fragmentation of a boosted $b\bar{b}$ pair (at hadronic machines like LHC or at $Z$ factories like LEP or FCC-$ee$), the time $t$ corresponds to the evolution between the production and the decay of a single neutral $B$-meson (ranging from $0$ to $+\infty$, with an exponential factor of the form $\exp(-\Gamma t)$).

The additional observables stemming from this interference between mixing and decay can in principle determine the phases of the Wilson coefficients $C_L^\nu$. Their derivation is a fairly straightforward extension of similar studies in the case of $b\to s\ell^+\ell^-$ modes~\cite{Descotes-Genon:2015hea,Descotes-Genon:2020tnz}. In the next subsections we outline how to adapt these results to the neutrino case and discuss these new observables.

The operators involved in the effective Hamiltonian  Eq.~(\ref{eq:hameffbsnunu}) can be recovered formally from the $b\to s\ell^+\ell^-$ Hamiltonian used in Refs.~\cite{Descotes-Genon:2015hea,Descotes-Genon:2020tnz} through the identification
\begin{equation}
C_9 \to C_L^\nu\,, \qquad
C_{10} \to - C_L^\nu\,, \qquad 
C_{9'} \to C_R^\nu\,, \qquad
C_{10'} \to - C_R^\nu\,, \qquad 
\end{equation}
where all the other (NP) Wilson coefficients vanish and a summation over neutrino flavours is required. We follow the notation of Ref.~\cite{Dunietz:2000cr} with
$\Delta m=M_H-M_L$ and $\Delta \Gamma=\Gamma_L-\Gamma_H$ where $L,H$ denote the lighter and heavier mesons. We have $x=\Delta m/\Gamma$ (0.77 for $B_d$, 26.8 for $B_s$) and $y=\Delta \Gamma/(2\Gamma)$ 
(0 for $B_d$, 0.07 for $B_s$)~\cite{ParticleDataGroup:2020ssz}.

\subsection{$B\to P\nu\bar\nu$}

First focusing on the $B_d\to K_S$ transition, adapting Refs.~\cite{Descotes-Genon:2015hea,Descotes-Genon:2020tnz} (the extension to $B_s\to f_0$ is immediate), we find that two amplitudes are involved to describe the $b\to s\nu\bar\nu$ transition:
\begin{eqnarray}
\bar{h}^\nu_V&\to& {\cal N}\frac{\sqrt{\lambda_B}}{2\sqrt{q^2}} (C_L^\nu+C_R^\nu) f_+ \,,\\
\bar{h}^\nu_A&\to& -{\cal N}\frac{\sqrt{\lambda_B}}{2\sqrt{q^2}} (C_L^\nu+C_R^\nu) f_+ \,,
\end{eqnarray}
where ${\cal N}$ is an overall normalisation factor proportional to $V_{tb}V_{ts}^*$ and $f_{+}$ is the vector form factor. The amplitudes $h$ can be obtained from $\bar{h}$ by applying the complex conjugation to both the normalisation factor ${\cal N}$ and the Wilson coefficients~\footnote{Contrary to $b\to s\ell\ell$, there are no CP-even (``strong'') phases to take into account from charm-loop contributions~\cite{Khodjamirian:2010vf,Khodjamirian:2012rm,Capdevila:2017ert,Bobeth:2017vxj,Gubernari:2020eft}.}.

An angular analysis based 
on the direction of the emitted leptons is impossible, and the only observable becomes (up to effects due to mixing)
\begin{equation}
\left.\frac{d\Gamma(\bar{B}_d\to K_S\nu\bar\nu)}{dq^2}\right|_{\rm no\ mixing}
 = 2\sum_\nu \bar{G}^\nu_0=2\sum_\nu \left(\frac{4}{3}\left[|\bar{h}^\nu_V|^2+|\bar{h}^\nu_A|^2\right]\right)\,,
\end{equation}
where $q^2$ is the dilepton invariant mass. Similar expressions can be obtained for the CP-conjugate mode.
We can study the impact of mixing through the time dependence of
$d\Gamma(B_d\to K_S\nu\bar\nu)/dq^2+d\Gamma(\bar{B}_d\to K_S\nu\bar\nu)/dq^2$ which is proportional to the sum over the neutrino flavours of
\begin{equation}
G^\nu_0(t)+\bar{G}^\nu_0(t)=e^{-\Gamma t} [(G^\nu_0+\bar{G}^\nu_0)\cosh(y\Gamma t)-h^\nu_0 \sinh(y\Gamma t)]\,.
\end{equation}
The term representing the interference between mixing and decay is~\footnote{We omit $\eta_M$ from our expressions derived from Ref.~\cite{Descotes-Genon:2015hea} since it equals 1 for both $K_S$ and $f_0$. The same will occur in the next secton for outgoing light vector mesons based on Ref.~\cite{Descotes-Genon:2020tnz} since $\eta_{M_1M_2}=1$ for all the cases considered.}
\begin{equation}
h^\nu_0=
{\rm Re}\left[e^{i\phi}  {\cal N}^2 (C_L^\nu+C_R^\nu)^2 \right] \frac{8}{3}\frac{\lambda_B(q^2)}{q^2}f_+^2(q^2)\,,
\end{equation}
with $\lambda=\lambda(m_B^2,m_P^2,q^2)$, so that $h_0$ is proportional to the interference ${\rm Re}[e^{i\phi} (V_{tb}V_{ts}^*)^2(C_L^\nu+C_R^\nu)^2]$. In the SM the neutral-meson mixing angle is $\phi=-2\beta$ for $B_d$ (and $\phi=2\beta_s$ for $B_s$)~\footnote{We will keep $\phi$ explicitly for generality in the following, although we will assume that there is no NP weak phase affecting $\Delta F=2$ transitions once we perform a numerical estimation of the observables.}.

The time-integrated version yields
\begin{equation}\label{eq:ratioBKSellell}
\frac{\langle d\Gamma(B_d\to K_S\nu\bar\nu)+d\Gamma(\bar{B}_d\to K_S\nu\bar\nu) \rangle_{\rm incoherent}}
{\langle d\Gamma(B_d\to K_S\nu\bar\nu)+d\Gamma(\bar{B}_d\to K_S\nu\bar\nu) \rangle_{\rm coherent}}
=1-y \frac{\sum_\nu h^\nu_0}{\sum_\nu (G^\nu_0+\bar{G}_0^\nu)}\,,
\end{equation}
where the brackets denote the integration over time. The coherent expression corresponds to the integration over $\Delta t$ at the B-factories (which yields a result identical to the case without mixing) while the incoherent expression denotes the integration over time $t$ between $B$ production and decay and is possibly applicable for the FCC-$ee$. We notice that the form factor $f_+$ dependence drops out from Eq.~(\ref{eq:ratioBKSellell}).

We can also consider the CP-asymmetry:
\begin{equation}\label{eq:CP-asymmetryABKS}
A^{\cal B}=\frac{d\Gamma(B_d\to K_S\nu\bar\nu)-d\Gamma(\bar{B}_d\to K_S\nu\bar\nu)}{d\Gamma(B_d\to K_S\nu\bar\nu)+d\Gamma(\bar{B}_d\to K_S\nu\bar\nu)}\,.
\end{equation}
In the above expression, the numerator is proportional to the sum over neutrino flavours of
\begin{equation}
G^\nu_0(t)-\bar{G}^\nu_0(t)=e^{-\Gamma t} [(G^\nu_0-\bar{G}^\nu_0)\cos(x\Gamma t)-s^\nu_0 \sin(x\Gamma t)]\,,
\end{equation}
with 
\begin{equation}
s^\nu_0=
{\rm Im}\left[e^{i\phi} {\cal N}^2 (C_L^\nu+C_R^\nu)^2 \right] \frac{8}{3}\frac{\lambda_B(q^2)}{q^2}f_+^2(q^2)\,,
\end{equation}
so that $s^\nu_0$ will be proportional to ${\rm Im}[e^{i\phi} (V_{tb}V_{ts}^*)^2(C_L^\nu+C_R^\nu)^2]$.

The time-integrated asymmetry vanishes for coherent production, but not for incoherent production where
\begin{equation}
\langle A^{\cal B}\rangle_{\rm incoherent}=\frac{1-y^2}{1+x^2} \frac{-x\sum_\nu s^\nu_0}{2\sum_\nu (G^\nu_0+\bar{G}^\nu_0-y h^\nu_0)}
\end{equation}
and we integrated the numerator and the denominator of the asymmetry $A^{\cal B}$ over time before taking the ratio.  This asymmetry in both time-dependent and time-integrated versions is free from hadronic uncertainties since the contribution from the form factor $f_+$ cancels in the ratio. 

Given the different values of $x$ and $y$ for the $B_{d,s}$ mesons, we see that
\begin{itemize}
\item for $B_d\to K_S\nu\bar\nu$, no information can be gathered on $h^\nu_0$, but $\sum_\nu s^\nu_0$ can be obtained from $A^{\cal B}$ either through a time-dependent analysis or through the time-integrated observable in incoherent production $\langle A^{\cal B}\rangle_{\rm incoherent}$, leading to a constraint on ${\rm Im}[e^{-2i\beta} (V_{tb}V_{ts}^*)^2(C_L^\nu+C_R^\nu)^2]$.
\item for $B_s\to f_0\nu\bar\nu$, one could in principle obtain information on both $\sum_\nu h^\nu_0$ and $\sum_\nu s^\nu_0$, but the challenging experimental identification of the $f_0$ meson and the current uncertainties in its theoretical description makes this decay very difficult to exploit.
\end{itemize}

\subsection{$B\to V\nu\bar\nu$}

In the case of $B_d\to K^{*0}(\to K_S \pi^0)\nu\bar\nu$ and $B_s\to\phi\nu\bar\nu$, we can adapt the description of the time-dependent observables for $b\to s\ell\ell$ given in Ref.~\cite{Descotes-Genon:2015hea}. We focus again on $B_d\to K^{*0}$ with an obvious extension to $B_s\to\phi$.
These decays are described by 8 transversity amplitudes for charged lepton modes~\cite{Altmannshofer:2009ma,Egede:2010zc}, out of which only 3 are non vanishing \footnote{The $A_t(q^2)$ amplitude as described in Ref.~\cite{Descotes-Genon:2015hea} does not vanish, but it is suppressed by the lepton mass and thus does not contribute here.} when describing the $b\to s\nu \bar \nu$ transition
\begin{eqnarray}
\bar{A}_\perp^{L} &\to&\bar{A}_\perp^{L\nu}= 2\sqrt{2}\sqrt{\lambda} \mathcal{N}'(C_L^\nu+C_R^\nu) \frac{V(q^2)}{m_B+m_{K^*}}\,, \\
\bar{A}_\parallel^L&\to&\bar{A}_\parallel^{L\nu}=  -2\sqrt{2}(m_B+m_{K^*})\mathcal{N}' (C_L^\nu-C_R^\nu) A_1(q^2)\,,   \\
\bar{A}_0^L&\to&\bar{A}_0^{L\nu}=  -\frac{(m_B+m_{K^*})\mathcal{N}'}{m_{K^*}\sqrt{q^2}} (C_L^\nu-C_R^\nu) \\
&&\times\left[(m_B^2-m_{K^*}^2-q^2)A_1(q^2)-\frac{\lambda}{(m_B+m_{K^*})^2}A_2(q^2)\right]\,,\nonumber
\end{eqnarray}
where the normalisation $\mathcal{N}'$ is proportional to $V_{tb}V_{ts}^*$ and $V$, $A_1$ and $A_2$ are the relevant hadronic form factors. The amplitudes $A$ are obtained from $\bar{A}$ by applying complex conjugation to both the normalisation factor $\mathcal{N}'$ and the Wilson coefficients.

As opposed to the pseudoscalar mode, one angle is still available to perform an angular analysis ($\theta_M$), describing the angle of the two final $K_S$ and $\pi^0$ mesons with respect to the line of flight of the $B_d$ meson in the vector-meson rest frame. After integrating over the azimuthal ($\phi$)  and the lepton polar ($\theta_\ell$) angles, the differential decay width in the absence of mixing takes the following form
\begin{equation}
   \left.\frac{d\Gamma(B_d\to K^{*0}\nu\bar\nu)}{dq^2d\cos\theta_M}\right|_{\rm no\ mixing}=\sum_\nu \left[\frac{3}{2}J^\nu_{1c}\cos^2\theta_M+J^\nu_{1s}\sin^2\theta_M\right]\,,
\end{equation}
where 
\begin{equation}
J^\nu_{1c}\to |A_0^{L\nu}|^2\,, \qquad J^\nu_{1s}\to \frac{3}{4}\left[|A_\perp^{L\nu}|^2+|A_\parallel^{L\nu}|^2\right]\,.
\end{equation}
In the absence of mixing, an angular analysis yields two observables, typically the branching fraction and the longitudinal $K^*$ polarization ($F_L$), proportional to $\sum_\nu (J^\nu_{1c}+\bar{J}^\nu_{1c}+J^\nu_{1s}+\bar{J}^\nu_{1s})$ and  $\sum_\nu (J^\nu_{1c}+\bar{J}^\nu_{1c})$ respectively. The CP-averaged decay rate and the CP-asymmetry can be considered in both cases, although the CP-asymmetry again vanishes due to the absence of strong phases, leading to the expressions in Eq.~(\ref{eq:BKnunu}) once evaluated.

In the presence of neutral-meson mixing, the time dependence of $d\Gamma(B_d\to K^{*0}(\to K_S\pi^0)\nu\bar\nu)+d\Gamma(\bar{B}_d\to K^{*0}(\to K_S\pi^0)\nu\bar\nu)$ is given by 
\begin{align}
    J^\nu_{i}(t)+\bar{J}^\nu_{i}(t)=e^{-\Gamma t}[(J^\nu_{i}+\bar{J}^\nu_{i})\cosh(y\Gamma t)-h^\nu_i\sinh(y\Gamma t)]\,,
\end{align}
whereas $d\Gamma(B_d\to K^{*0}(\to K_S\pi^0)\nu\bar\nu)-d\Gamma(\bar{B}_d\to K^{*0}(\to K_S\pi^0)\nu\bar\nu)$ involves 
\begin{align}
    J^\nu_{i}(t)-\bar{J}^\nu_{i}(t)=e^{-\Gamma t}[(J^\nu_{i}-\bar{J}^\nu_{i})\cos(x\Gamma t)-s^\nu_i\sin(x\Gamma t)]\,.
\end{align}

From these two sets of new observables, the simpler one corresponds to $\sum_\nu s^\nu_{1c}$ and $\sum_\nu h^\nu_{1c}$ which only features the longitudinal amplitude, explicitly
\begin{align}
\begin{split}
h^\nu_{1c}&=2{\rm Re}\left[e^{i\phi}(\bar{A}_0^{L\nu})^2\right]=2 |A_0^{L\nu}|^2 \frac{{\rm Re}[e^{i\phi}(V_{tb}V_{ts}^*)^2 (C_L^{\nu}-C_R^{\nu})^2]}{|V_{tb}V_{ts}^*|^2 |C_L^\nu-C_R^\nu|^2}\,,
\end{split}\\
\begin{split}
s^\nu_{1c}&=2{\rm Im}\left[e^{i\phi}(\bar{A}_0^{L\nu})^2\right]=2 |A_0^{L\nu}|^2 \frac{{\rm Im}[e^{i\phi}(V_{tb}V_{ts}^*)^2 (C_L^{\nu}-C_R^{\nu})^2]}{|V_{tb}V_{ts}^*|^2 |C_L^\nu-C_R^\nu|^2}  \,,
\end{split}  
\end{align}
and thus contain the same amplitude as the numerator of $F_L$.
In the case of $h^\nu_{1s}$ and $s^\nu_{1s}$ they involve different amplitudes and combinations of the Wilson coefficients
\begin{align}
\begin{split}
h^\nu_{1s}&=\frac{3}{2}{\rm Re}\left[e^{i\phi}\left\{(\bar{A}_\parallel^{L\nu})^2-(\bar{A}_\perp^{L\nu})^2\right\}\right]\\
&=12 |{\cal N'}|^2 (m_B+m_{K^{*0}})^2 [A_1(q^2)]^2 {\rm Re}[Z^\nu]\,,\end{split}\\
\begin{split}
s^\nu_{1s}&=\frac{3}{2}{\rm Im}\left[e^{i\phi}\left\{(\bar{A}_\parallel^{L\nu})^2-(\bar{A}_\perp^{L\nu})^2\right\}\right]
\\&=12 |{\cal N'}|^2 (m_B+m_{K^{*0}})^2 [A_1(q^2)]^2 {\rm Im}[Z^\nu]\,,\end{split}\\
\begin{split}
Z^\nu&=e^{i\phi}\frac{(V_{tb}V_{ts}^*)^2}{|V_{tb}V_{ts}^*|^2 }\left( (C_L^{\nu}-C_R^{\nu})^2+(C_L^{\nu}+C_R^{\nu})^2\frac{\lambda_B(q^2) [V(q^2)]^2}{(m_B^2+m_{K^{*0}}^2)^2A_1(q^2)^2}\right)\,,
\end{split}
\end{align}
 making them less clean observables since the dependence on the form factors does not drop out from ratios.

The time-integrated observables are similar to the pseudoscalar case, yielding the CP-averaged observables
\begin{equation}
\frac{\langle \cal B\rangle_{\rm incoherent}}{\langle \cal B\rangle_{\rm coherent}}=1-y \frac{\sum_\nu (h^\nu_{1c}+\frac{4}{3}h^\nu_{1s})}{\sum_\nu [(J^\nu_{1c}+\bar{J}^\nu_{1c})+\frac{4}{3}(J^\nu_{1s}+\bar{J}^\nu_{1s})]}\,,
\end{equation}
\begin{equation}
\frac{\langle F_L\rangle_{\rm incoherent}}{\langle F_L\rangle_{\rm coherent}}\frac{\langle \cal B\rangle_{\rm incoherent}}{\langle \cal B\rangle_{\rm coherent}}=1-y\frac{\sum_\nu h^\nu_{1c}}{\sum_\nu (J^\nu_{1c}+\bar{J}^\nu_{1c})}\,,
\end{equation}
and the CP-asymmetries defined similarly to Eq.~(\ref{eq:CP-asymmetryABKS}):
\begin{eqnarray}
\!\!\!\!\!\!\!\!\!\!&&\langle A^{\cal B}\rangle _{\rm incoherent}=\\ \nonumber
\!\!\!\!\!\!\!\!\!\!&&\qquad \frac{1-y^2}{1+x^2} \frac{-x\sum_\nu(s_{1c}+\frac{4}{3}s^\nu_{1s})}{2\sum_\nu[(J^\nu_{1c}+\bar{J}^\nu_{1c})+\frac{4}{3}(J^\nu_{1s}+\bar{J}^\nu_{1s})-y(h^\nu_{1c}+\frac{4}{3}h^\nu_{1s})]}\,,\\
\!\!\!\!\!\!\!\!\!\!&&\langle A^{F_L}\rangle _{\rm incoherent}=\\\nonumber
\!\!\!\!\!\!\!\!\!\!&&\qquad \frac{1-y^2}{1+x^2} \frac{-x\sum_\nu s^\nu_{1c}}{2\sum_\nu[(J^\nu_{1c}+\bar{J}^\nu_{1c})+\frac{4}{3}(J^\nu_{1s}+\bar{J}^\nu_{1s})-y(h^\nu_{1c}+\frac{4}{3}h^\nu_{1s})]}\,,
\end{eqnarray}
which can be combined into clean ratios without hadronic uncertainties
\begin{eqnarray}
\!\!\!\!\!\!\!\!\!\!&&\frac{\langle A^{F_L}\rangle_{\rm incoherent}}{\langle F_L \rangle_{\rm incoherent}}=\frac{1-y^2}{1+x^2} \frac{-x \sum_\nu s_{1c}}{ \sum_\nu (J_{1c}+\bar{J}_{1c}-y \, h_{1c})}\nonumber \\ \nonumber
\!\!\!\!\!\!\!\!\!\!&&=\frac{1-y^2}{1+x^2} \frac{-x\ {\rm Im}[e^{i\phi}(V_{tb}V_{ts}^*)^2  \sum_\nu(C_L^{\nu}-C_R^{\nu})^2]}{
 \sum_\nu \{ |V_{tb}V_{ts}^*|^2 |C_L^\nu-C_R^\nu|^2-y{\rm Re}[e^{i\phi}(V_{tb}V_{ts}^*)^2 (C_L^{\nu}-C_R^{\nu})^2] \} }
\nonumber\\
\!\!\!\!\!\!\!\!\!\!&&=\frac{1-y^2}{1+x^2}\times \frac{-x\sum_\nu \sin(\phi-2\beta_s+2\alpha^\nu_{LR})}{1-y\sum_\nu \cos(\phi-2\beta_s+2\alpha^\nu_{LR})}\,,
\end{eqnarray}
where $\alpha^\nu_{LR}$ is the argument of $C_L^{\nu}-C_R^{\nu}$.

Given the different values of the mixing parameters for the two neutral mesons, we see that
\begin{itemize}
\item for $B_d\to K^*(\to K_S\pi^0)\nu\bar\nu$,  the only relevant effects are on the CP asymetries through  $A^{F_L}$ (or equivalently $\sum_\nu s^\nu_{1c}$) constraining ${\rm Im}[e^{-2i\beta}(V_{tb}V_{ts}^*)^2 (C_L^{\nu}-C_R^{\nu})^2]$
and through $A^{\cal B}$ (or equivalently $\sum_\nu 3s^\nu_{1c}+4s^\nu_{1s}$)
which corresponds to  a  less clean combination of ${\rm Im}[e^{-2i\beta}(V_{tb}V_{ts}^*)^2\sum_\nu(C_L^\nu-C_R^\nu)^2]$ and ${\rm Im}[e^{-2i\beta}(V_{tb}V_{ts}^*)^2\sum_\nu(C_L^\nu+C_R^\nu)^2]$, involving form factors.
\item for $B_s\to \phi\nu\bar\nu$,  CP-averaged quantities are also accessible, so that $F_L$ could yield a constraint on ${\rm Re}[\sum_\nu(C_L^\nu-C_R^\nu)^2]$, while $\cal B$ would yield a less clean constrain on a combination of ${\rm Re}[\sum_\nu(C_L^\nu-C_R^\nu)^2]$ and ${\rm Re}[\sum_\nu(C_L^\nu+C_R^\nu)^2]$, which depends on the form factors.
\end{itemize}
This above discussion assumes that there is no NP in mixing, but it can be easily extended if there is an addition NP phase contributing to $\phi$.

\subsection{Summary}

If we consider a $b\to s\nu\bar\nu$ transition corresponding to a $B$-meson decaying into a meson $M$, we can get new observables by considering the time dependence of CP-asymmetries (for $B_d\to M\nu\bar\nu$ and $B_s\to M\nu\bar\nu$) and CP-averaged observables (for $B_s\to M\nu\bar\nu$ only). These can be studied through the time dependence after a coherent $b\bar{b}$ production at B-factories, but also through time-integrated observables after an incoherent $b\bar{b}$ production at hadronic machines or Z-factories : the CP-asymmetries are non-vanishing (contrary to B factories) whereas the CP-averaged quantities deviate from the B-factory values. In all cases, one can build ratios of angular observables where form factor dependencies cancel, providing unique clean probes of NP phases in the Wilson coefficients responsible for CP-violation in $b\to s\nu\bar\nu$, even in absence of right-handed currents.


\section{Numerical study of NP sensitivity and projections}\label{sec:numerical}

In the following we present a brief numerical study, to illustrate the possible size of NP effects in the observables discussed above and the sensitivity that could be potentially reached by future experiments. 

\subsection{NP benchmark} \label{sec:NPBenchmark}

For illustration purposes, we consider a subset of relevant SMEFT operators, considered previously in Ref.~\cite{Descotes-Genon:2020buf}, focusing on the case where NP involves only left-handed fields. Possible heavy NP contributions should be written in terms of SM gauge invariant operators~\cite{DAmbrosio:2002vsn,Hurth:2008jc,Buttazzo:2017ixm}:
\begin{align}\label{eq:ops}
& \mathcal L_{\rm eff.}   = \mathcal L_{\rm SM}  - \frac{1}{v^2} \lambda^q_{ij} \lambda^\ell_{\alpha\beta}  \\   & \quad\times
\Big[ C_T \left( \bar Q_L^i \gamma_\mu \sigma^a Q_L^i \right) \left( \bar L_L^\alpha \gamma^\mu \sigma^a L_L^\beta \right)    +  C_S \left( \bar Q_L^i \gamma_\mu  Q_L^i \right) \left( \bar L_L^\alpha \gamma^\mu  L_L^\beta \right)\Big], \nonumber 
\end{align}
where we choose to write the operators in the down-quark and charged-lepton mass basis
$Q_L^i = (V^{\rm CKM *}_{ji}  u_L^j, d_L^i)^T$
and $L_L^\alpha = (U^{\rm PMNS}_{\alpha\beta}\nu_L^\beta,\ell^{\alpha}_L)^T$. Following Refs.~\cite{Buttazzo:2017ixm,Bordone:2017lsy,Cornella:2019hct}
we assume that the same flavour structure encoded in (hermitian) $\lambda^q_{ij}$ and $\lambda^\ell_{\alpha\beta}$ matrices holds for both operators, while $C_{S,T}$ are real.


For the quarks only $\lambda^q_{23,32}$ elements contribute to $b\to s$ transitions in this basis. In the lepton sector, we consider here only (SM-like) left-handed neutrinos and in addition assume an approximate $U(1)^3_\ell$ symmetry (broken only by the neutrino masses). This  yields $\lambda_{i\neq j}^\ell \simeq 0$\, in agreement with the stringent experimental limits on lepton flavour violation.   Current (LFU) NP hints in $b\to s\ell\ell$ decays only indicate significant non-standard effects in muonic final states, whereas a slightly smaller effect in electrons is not excluded. Tauonic transitions are at present only poorly constrained and could in principle exhibit even much larger deviations than those observed in $R_{K^{(*)}}$~\cite{Capdevila:2017iqn}.  We may thus assume various ratios of $U(1)_\ell^3$ charges~\cite{Descotes-Genon:2020buf}: 1) the simplest case $\lambda^{\ell}_{ee} =\lambda^{\ell}_{\tau\tau} = 0$; 2) the democratic scenario $\lambda^{\ell}_{ee} =\lambda^{\ell}_{\mu\mu}=\lambda^{\ell}_{\tau\tau}$; 3) the anomaly-free assignment $\lambda^{\ell}_{\mu\mu} = -\lambda^{\ell}_{\tau\tau}$ and $\lambda^{\ell}_{ee} =0$; 4) the hierarchical charge scenario $\lambda^{\ell}_{\alpha\alpha} / \lambda^{\ell}_{\mu\mu} = m_\alpha / m_\mu$.

It is easy to work out the values of the Wilson coefficients for the effective theories~\cite{Descotes-Genon:2020buf} for $b\to s\ell\ell$ and $b\to s\nu\bar\nu$:
\begin{eqnarray}\label{eq:C9mu}
C^{\mu,{\rm NP}}_{9} &=  &- \frac{\pi}{\alpha_{em} V_{tb}V_{ts}^* } \lambda_{23}^q \lambda_{\mu\mu}^\ell [C_T +C_S]
\,,\\
C^{\mu,{\rm NP}}_{10}&=  &- \frac{\pi}{\alpha_{em} V_{tb}V_{ts}^* } \lambda_{23}^q \lambda_{\mu\mu}^\ell  [- C_T -C_S]  \,, 
\end{eqnarray}
wheras other $b\to s\ell\ell$ Wilson coefficients do not receive any NP contributions,  and
\begin{align}
C^{\nu_\alpha, {\rm NP}}_L &=   -\frac{\pi}{\alpha_{em} V_{tb}V_{ts}^*}\lambda_{23}^q \lambda_{\alpha\alpha}^\ell [C_S-C_T]\,, \label{eq:CnuL} \\
 C^{\nu_\alpha, {\rm NP}}_R &= 0 \label{eq:CnuR}\,,
 \end{align}
 with $\alpha=e,\mu,\tau$.\footnote{We neglect tiny neutrino mass effects, setting effectively $U_{\alpha\beta}^{\rm PMNS} = \delta^{\alpha\beta}$ and identifying neutrinos by the corresponding charged lepton flavor.}
Notice that in this setup, the three neutrino flavours receive a NP contribution $C^{\nu_\alpha, {\rm NP}}_L$ with the same CP-violating phase relative to the SM $\phi_{\rm NP} = {\rm arg} (\lambda^q_{23} / V_{tb}V_{ts}^*)$, whereas $C^{\nu_\alpha, {\rm NP}}_R$ vanishes~\footnote{Similar common flavour structures (and weak phase) naturally arise also in specifci flavour models, for instance those based on minimally broken $U(2)$ flavour symmetry respecting General Minimal Flavour Violation~\cite{Kagan:2009bn}, as can be seen from the discussion in Ref.~\cite{Descotes-Genon:2020buf}.} .
  
 We are now in a position to define our  benchmark NP model for $b\to s\nu\bar\nu$ .
 Global fits to $b\to s\ell^+\ell^-$ data suggest a value of $C_9^{\mu, {\rm NP}}$ around a quarter of its SM value for the scenario $C_9^{\mu, {\rm NP}}=-C_{10}^{\mu, {\rm NP}}$~\cite{Alguero:2021anc}. In Eqs.~(\ref{eq:C9mu})-(\ref{eq:CnuR}), $C^{\mu,{\rm NP}}_{9,10}$ and $C^{\nu_\mu, {\rm NP}}_{L,R}$ involved $C_S\pm C_T$ multiplied by the same normalisation. Therefore, a natural benchmark point consists in taking $C^{\nu_\alpha, {\rm NP}}_L  = \abs{C^{\nu_\alpha, {\rm NP}}_L } \exp(i\phi_{\rm NP})$ with $\phi_{\rm NP}$ arbitrarily large, while  $\abs{C^{\nu_\mu, {\rm NP}}_L }=C^{\nu, {\rm SM}}_L/4$, and rescaled according to the different scenarios for the $\lambda^\ell$ couplings in order to obtain $\abs{C^{\nu_e, {\rm NP}}_L}$ and $\abs{C^{\nu_\tau, {\rm NP}}_L}$.

Since $C^{\nu, {\rm NP}}_R$ vanishes for all three neutrino flavours, 
the usual observables discussed in Sec.~\ref{sec:btosnunu} cannot probe the phase $\phi_{\rm NP}$, but they are still sensitive to the moduli of $C^{\nu, {\rm NP}}_L$. We checked that the four scenarios considered for the lepton couplings are compatible with the current experimental bounds on the branching ratios for all values of $\phi_{\rm NP}$, and we will use these scenarios to illustrate the effect of the phase $\phi_{\rm NP}$ over the observables discussed in \cref{sec:probingNPweakphases}.

\begin{table}
\renewcommand{\arraystretch}{1.2}
\centering
\begin{tabular}{|c|cccc|}
\toprule
\multicolumn{5}{c}{$B\to K_S\nu\bar\nu$} \\
\midrule
\multirow{2}{*}{Obs}&\multirow{2}{*}{SM}&\multicolumn{3}{c|}{Exp Uncertainty }\\
                & &200  & 2000  & 20000 \\  \midrule
$\langle A^{\cal B}\rangle_{\rm incoherent}$& -0.356$\pm$ 0.005& 0.06 & 0.021 & 0.006 \\ 
$s_0/(\Gamma + \bar\Gamma )$&                           0.736$\pm$ 0.009          & 0.16 & 0.05 & 0.017 \\ \bottomrule 
\toprule
\multicolumn{5}{c}{$B\to K^*\nu\bar\nu$} \\
\midrule
\multirow{2}{*}{Obs}&\multirow{2}{*}{SM}&\multicolumn{3}{c|}{Exp Uncertainty }\\
                & &200  & 2000  & 20000 \\  \midrule
$\langle A^\mathcal{B}\rangle_{\rm incoherent}$& 0.257$\pm$ 0.010 & 0.07 & 0.022 & 0.007 \\ 
$\langle F_L\rangle_{\rm incoherent}$&   0.49$\pm$0.04   & 0.05 & 0.018 & 0.006 \\ 
$\langle A_{F_L}\rangle_{\rm incoherent}$&  0.173$\pm$0.014  & 0.06 & 0.021 & 0.006 \\ 
$\frac{\langle A_{F_L}\rangle_{\rm incoherent}}{\langle F_L\rangle_{\rm incoherent}}$& 0.356$\pm$0.005& 0.14 & 0.04 & 0.013  \\ 
$(J_{1c}+\bar{J}_{1c})/{(\Gamma + \bar\Gamma )}$&         0.49$\pm$0.04            & 0.13 & 0.04  & 0.011 \\  
$(J_{1c}-\bar{J}_{1c})/{(\Gamma + \bar\Gamma )}$&             0        & 0.15 & 0.04  & 0.011 \\  
$(J_{1s}+\bar{J}_{1s})/{(\Gamma + \bar\Gamma )}$&           0.386$\pm$0.029                  & 0.09 & 0.029 & 0.009 \\  
$(J_{1s}-\bar{J}_{1s})/{(\Gamma + \bar\Gamma )}$&                0          & 0.11 & 0.03  & 0.009 \\  
$s_{1c}/(\Gamma + \bar\Gamma )$&            -0.358$\pm$0.029                      & 0.14 & 0.05  & 0.015 \\  
$s_{1s}/(\Gamma + \bar\Gamma )$&           -0.130$\pm$0.018          & 0.11 & 0.04 & 0.013 \\  
$s_{1c}/(J_{1c}+\bar{J}_{1c})$&                  -0.736$\pm$0.009         & 0.4  & 0.13  & 0.04  \\  
$s_{1s}/(J_{1s}+\bar{J}_{1s})$&               -0.338$\pm$ 0.034            & 0.3 & 0.1  & 0.034 \\ 
\bottomrule
\toprule
\multicolumn{5}{c}{$B_s\to \phi\nu\bar\nu$} \\
\midrule
\multirow{2}{*}{Obs}&\multirow{2}{*}{SM}&\multicolumn{3}{c|}{Exp Uncertainty }\\
                & &200  & 2000  & 20000 \\  \midrule
$\langle A^{\cal B}\rangle_{\rm incoherent}$& 0.0& 0.07 & 0.023 & 0.007 \\ 
$\langle F_L\rangle_{\rm incoherent}$ &  0.570$\pm$0.021 & 0.05 & 0.017 & 0.005 \\ 
$\langle A_{F_L}\rangle_{\rm incoherent}$& 0 & 0.07 & 0.022 & 0.007 \\ 
$\frac{\langle A_{F_L}\rangle_{\rm incoherent}}{\langle F_L\rangle_{\rm incoherent}}$& 0& 0.12 & 0.04 & 0.012  \\ 
$(J_{1c}+\bar{J}_{1c})/{(\Gamma + \bar\Gamma )}$&              0.660$\pm$0.025       & 0.12 & 0.04  & 0.012 \\  
$(J_{1c}-\bar{J}_{1c})/{(\Gamma + \bar\Gamma )}$&             0        & 0.14 & 0.04  & 0.013 \\  
$(J_{1s}+\bar{J}_{1s})/{(\Gamma + \bar\Gamma )}$&             0.364$\pm$0.018        & 0.08 & 0.026 & 0.008 \\  
$(J_{1s}-\bar{J}_{1s})/{(\Gamma + \bar\Gamma )}$&            0         & 0.09 & 0.025  & 0.009 \\  
$s_{1c}/(\Gamma + \bar\Gamma )$&                  0.0$\pm$0.0008                  & 0.12 & 0.04  & 0.013 \\  
$s_{1s}/(\Gamma + \bar\Gamma )$&                  0.0$\pm$0.00035                   & 0.08 & 0.025 & 0.008 \\  
$h_{1c}/(\Gamma + \bar\Gamma )$&                0.66$\pm$0.025                    & 1.5 & 0.5  & 0.16 \\  
$h_{1s}/(\Gamma + \bar\Gamma )$&                   0.31$\pm$0.02                  & 1.1 & 0.34 & 0.12 \\  
$s_{1c}/(J_{1c}+\bar{J}_{1c})$&               0$\pm$0            & 0.2  & 0.06  & 0.02  \\  
$s_{1s}/(J_{1s}+\bar{J}_{1s})$&               0.000$\pm$0.001                  & 0.23 & 0.07  & 0.023 \\
$h_{1c}/(J_{1c}+\bar{J}_{1c})$&            1$\pm$0           & 2.4  & 0.8  & 0.26  \\  
$h_{1s}/(J_{1s}+\bar{J}_{1s})$&                 0.844$\pm$0.027          & 2.8 & 0.9  & 0.31 \\  
\bottomrule
\end{tabular}
\caption{SM predictions and experimental projections for the observables discussed in \cref{sec:probingNPweakphases} for the $B^0\to K_S\nu\bar\nu$, $B^0\to K^{*0}\nu\bar\nu$ and $B_s^0\to \phi\nu\bar\nu$ modes. All the observables shown are summed over the 3 neutrino flavours and integrated over the whole kinematic range. On the first column the SM predictions are shown while on the 3 following columns, the expected statistical uncertainty for 200, 2000 and 20000 events of the respective modes are given as discussed in \cref{sec:prospects}. }
\label{tab:projections}
\end{table}


\subsection{Prospects at Belle II, CEPC and FCC-$ee$}\label{sec:prospects}

Given the current sensitivity projections at Belle II and FCC-$ee$ discussed in Sec.~\ref{sec:introduction}, as conservative/optimistic scenarios, we make projections for 200, 2k and 20k reconstructed $B_d\to K_S\nu\bar\nu$, $B^0 \to K^{*0} \nu \bar \nu$  and $B_s \to \phi \nu \bar \nu$ decays. Our objective is to obtain a rough estimate of the projected statistical uncertainties for the time-dependent and time-integrated observables discussed before (we consider only observables integrated over the whole $q^2$ range in the following). 

\begin{figure}[t!]
\includegraphics[width=\columnwidth]{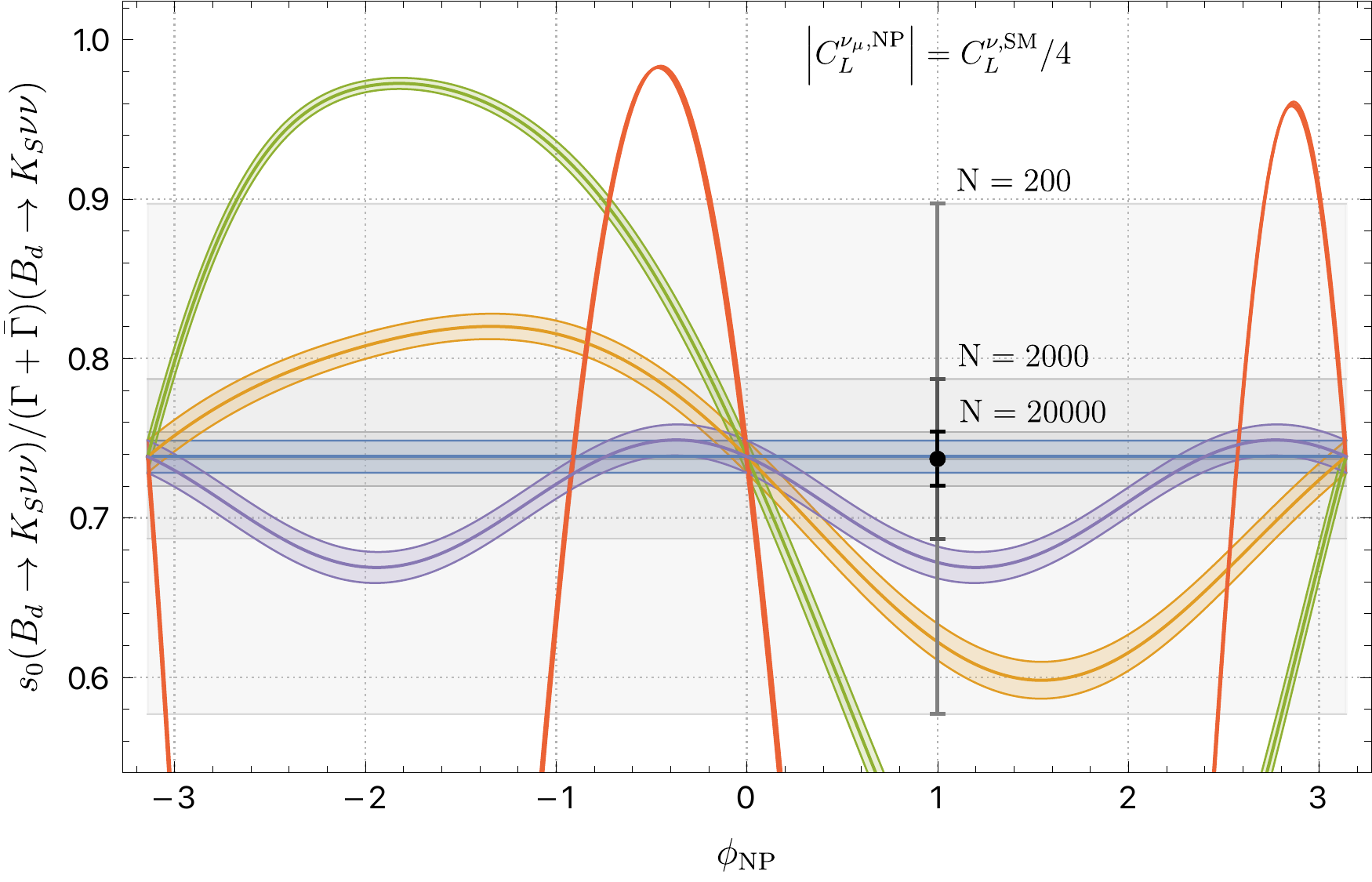}
\includegraphics[width=\columnwidth]{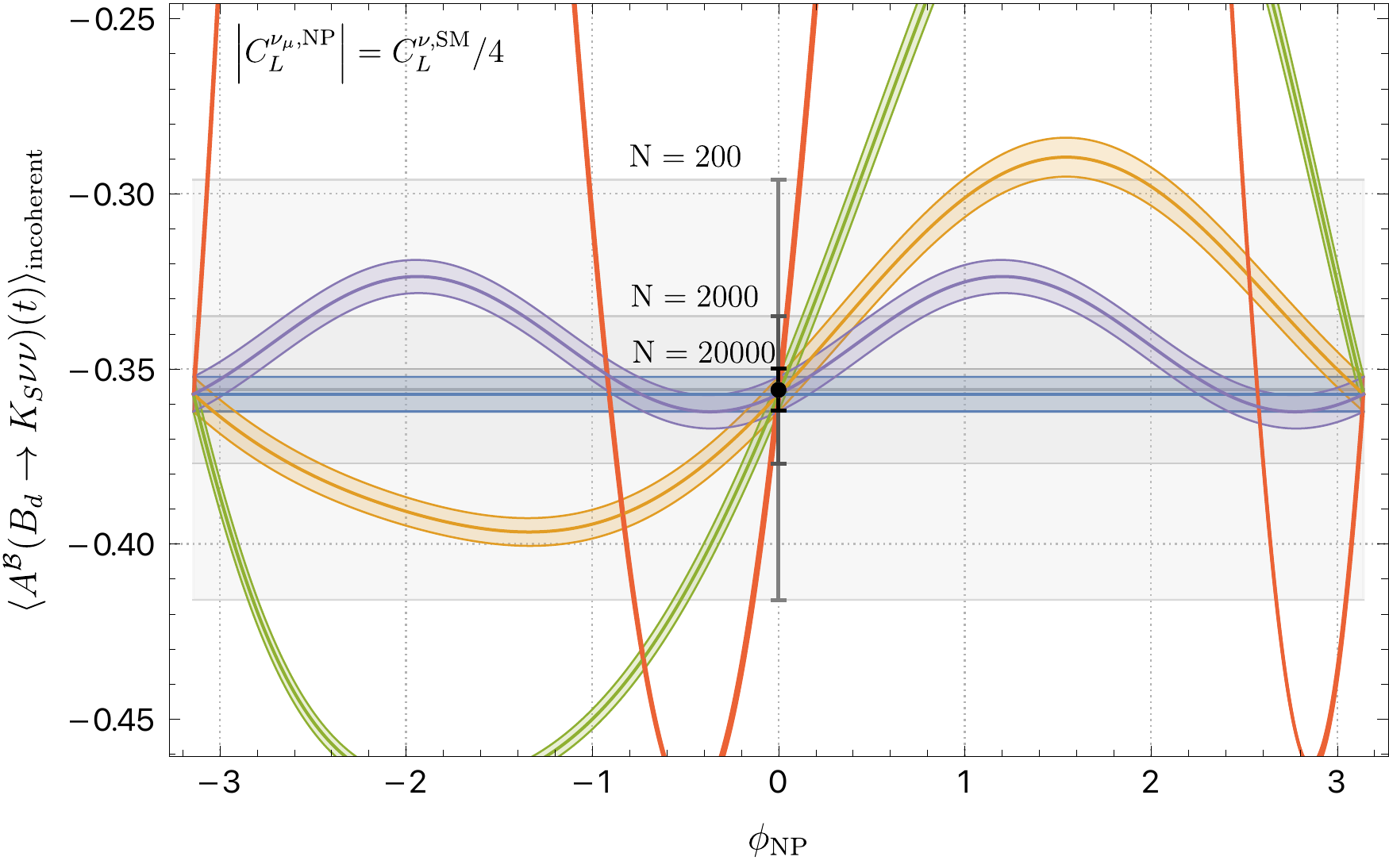}
\caption{
$\langle A^{\rm CP}\rangle_{\rm incoherent}$ (bottom) and
$s_0/(\Gamma + \bar\Gamma )$ (top) for the $B\to K_S\nu\bar\nu$ decay as a function of the complex phase $\phi_{\rm NP}$ of the NP Wilson coefficient $C_L^{\nu_\mu, {\rm NP}}=e^{-i\phi_{\rm NP}}\abs{C_L^{\nu_\mu, {\rm NP}}}$ for $\abs{C_L^{\nu_\mu, {\rm NP}}}=\abs{C_L^{\nu, {\rm SM}}}/4$ ($C_R^{\nu, {\rm NP}}=0$ is assumed). The NP Wilson coefficients for the other lepton flavours carry the same phase but are rescaled according to 
scenarios with the 1) simplest,  2)  democratic, 3) anomaly-free assignment, 4) hierarchical lepton structures described in \cref{sec:NPBenchmark}, leading to the variations shown in yellow, green, purple and orange respectively. 
 The SM prediction is shown in blue.
 The 3 gray bands correspond, from the widest to the narrrowest, to the expected experimental uncertainties given in \cref{tab:projections} for N=200, N=2000, and N=20000 events.  
}\label{fig:PlotsKS}
\end{figure}

\begin{figure}[t!]
\includegraphics[width=\columnwidth]{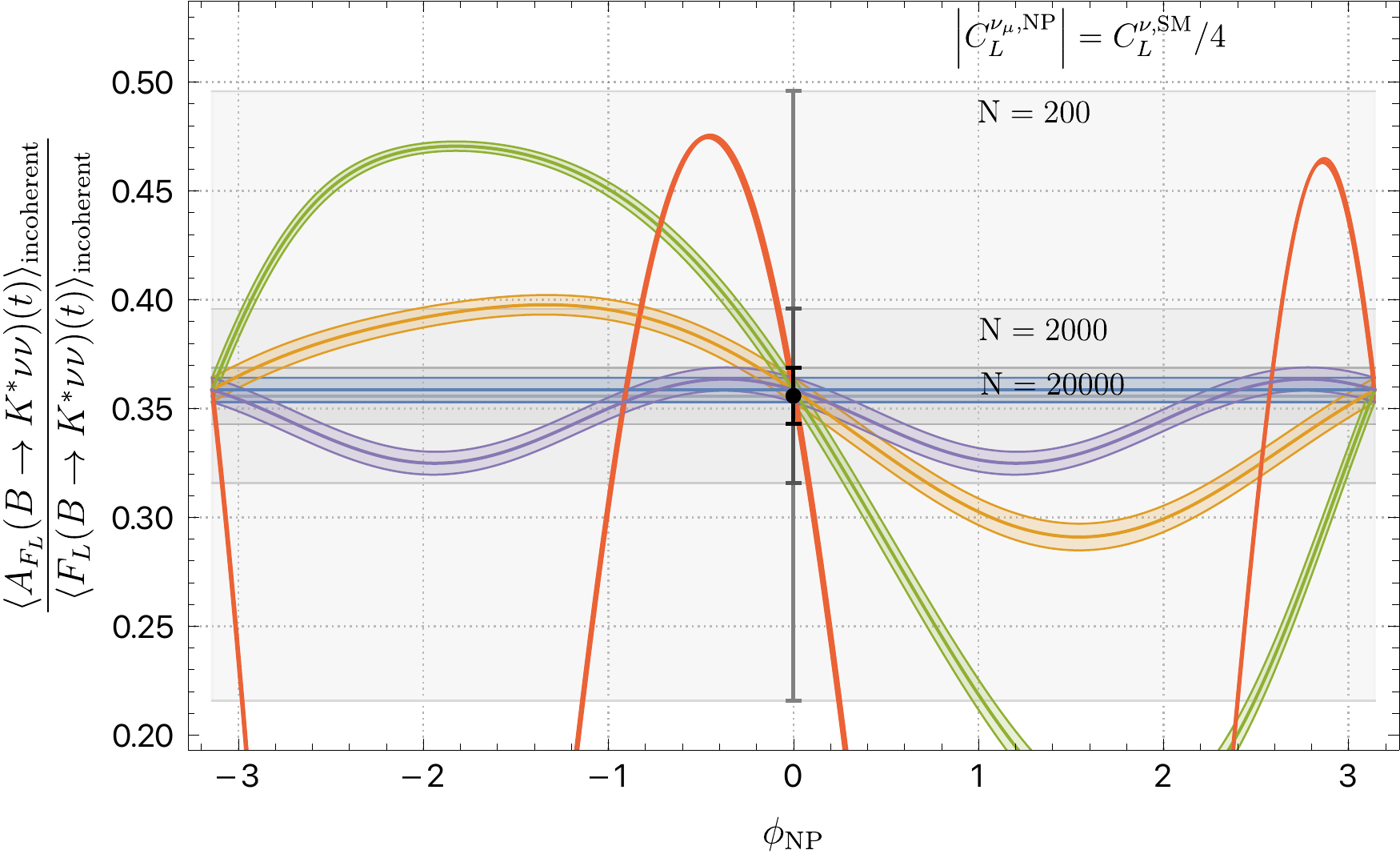}
\includegraphics[width=\columnwidth]{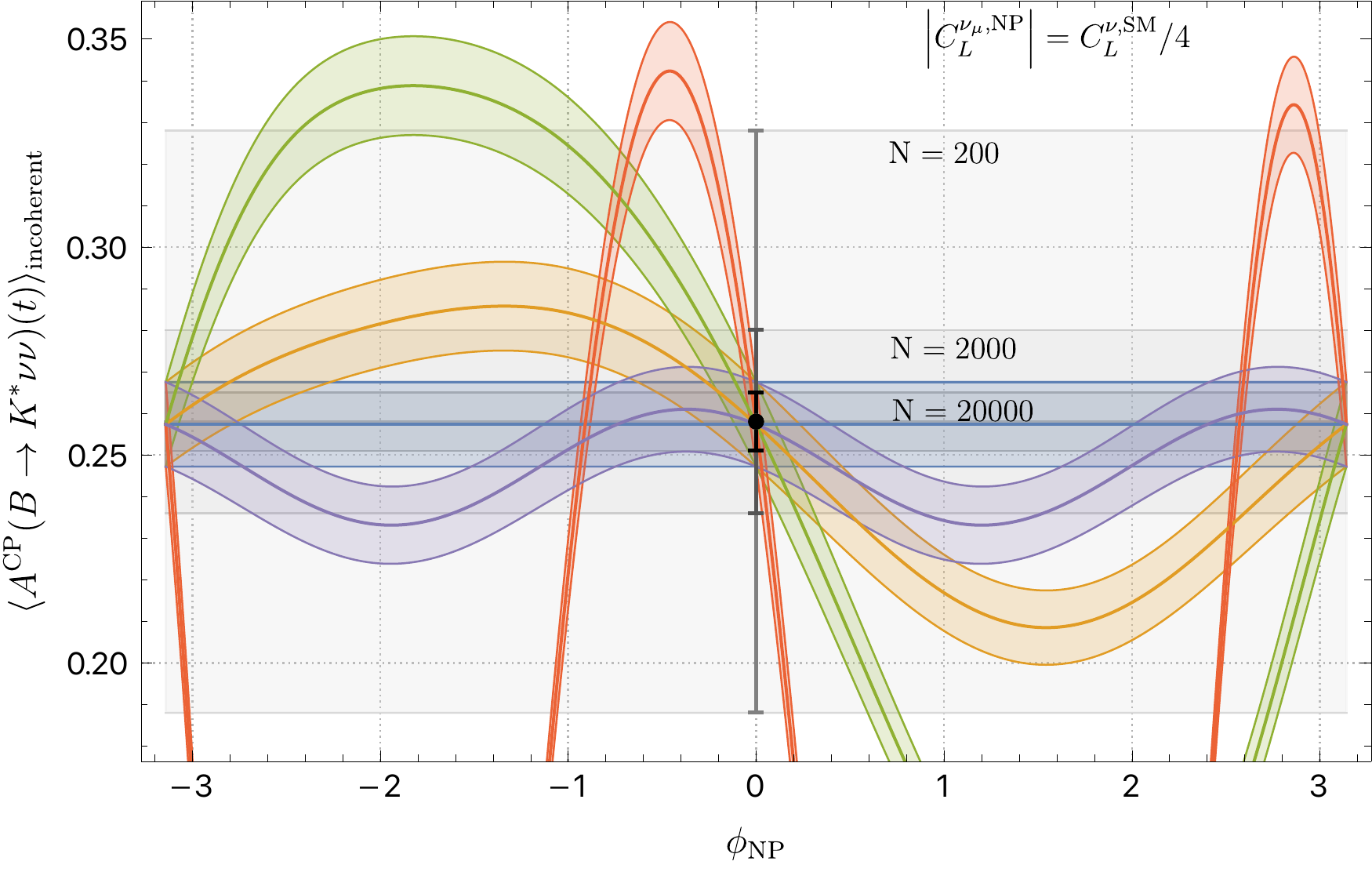}
\includegraphics[width=\columnwidth]{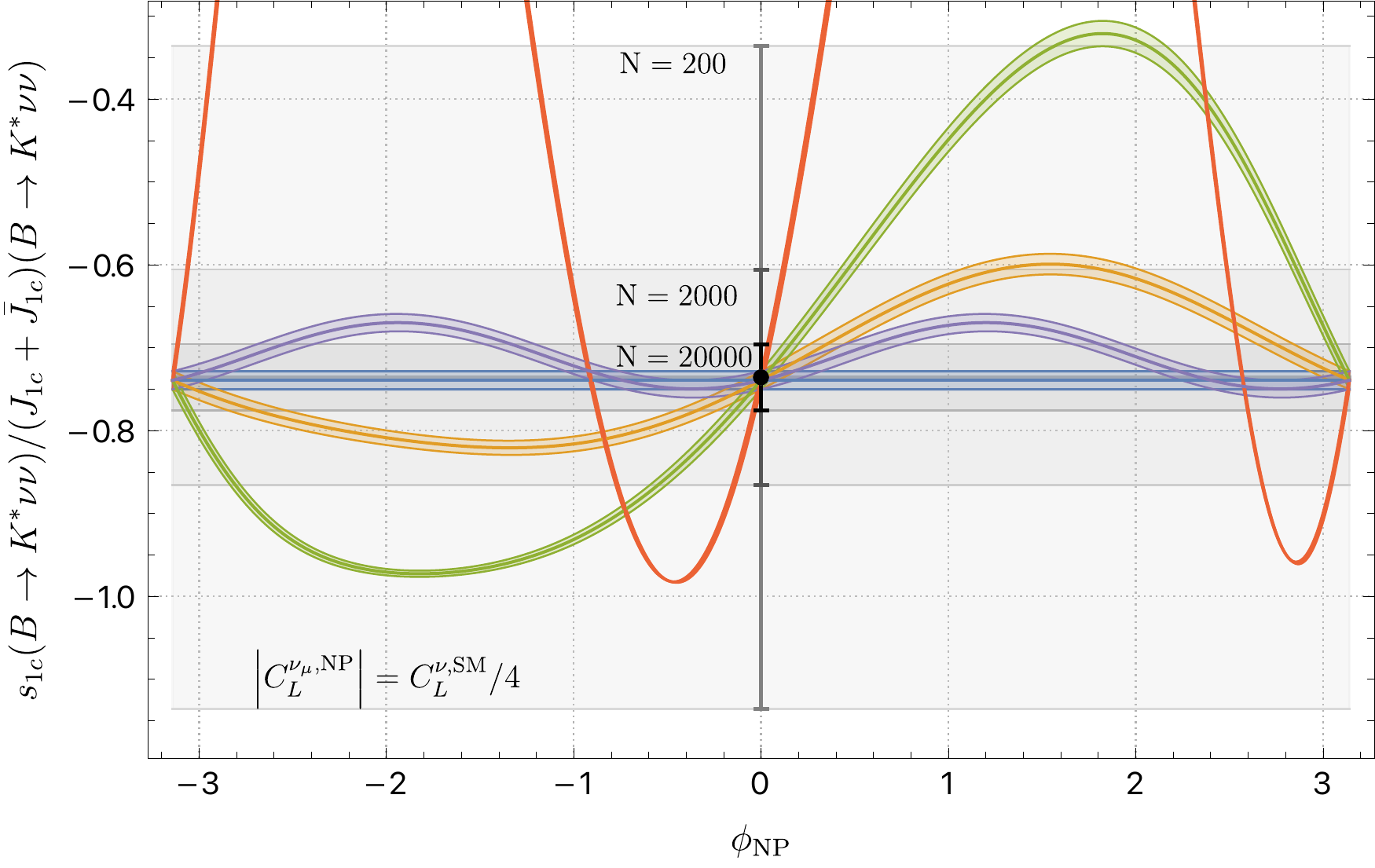}
\caption{Same as \cref{fig:PlotsKS} for the $\langle A_{F_L}\rangle_{\rm incoherent}/\langle F_L\rangle_{\rm incoherent}$  (top) $\langle A^{\cal B}\rangle_{\rm incoherent}$ (center) and
$s_{1c}/(J_{1c}+\bar{J}_{1c})$ (bottom) observables of the $B\to K^*\nu\bar\nu$ mode.}\label{fig:PlotsKstar}
\end{figure}

\begin{figure}[t!]
\includegraphics[width=\columnwidth]{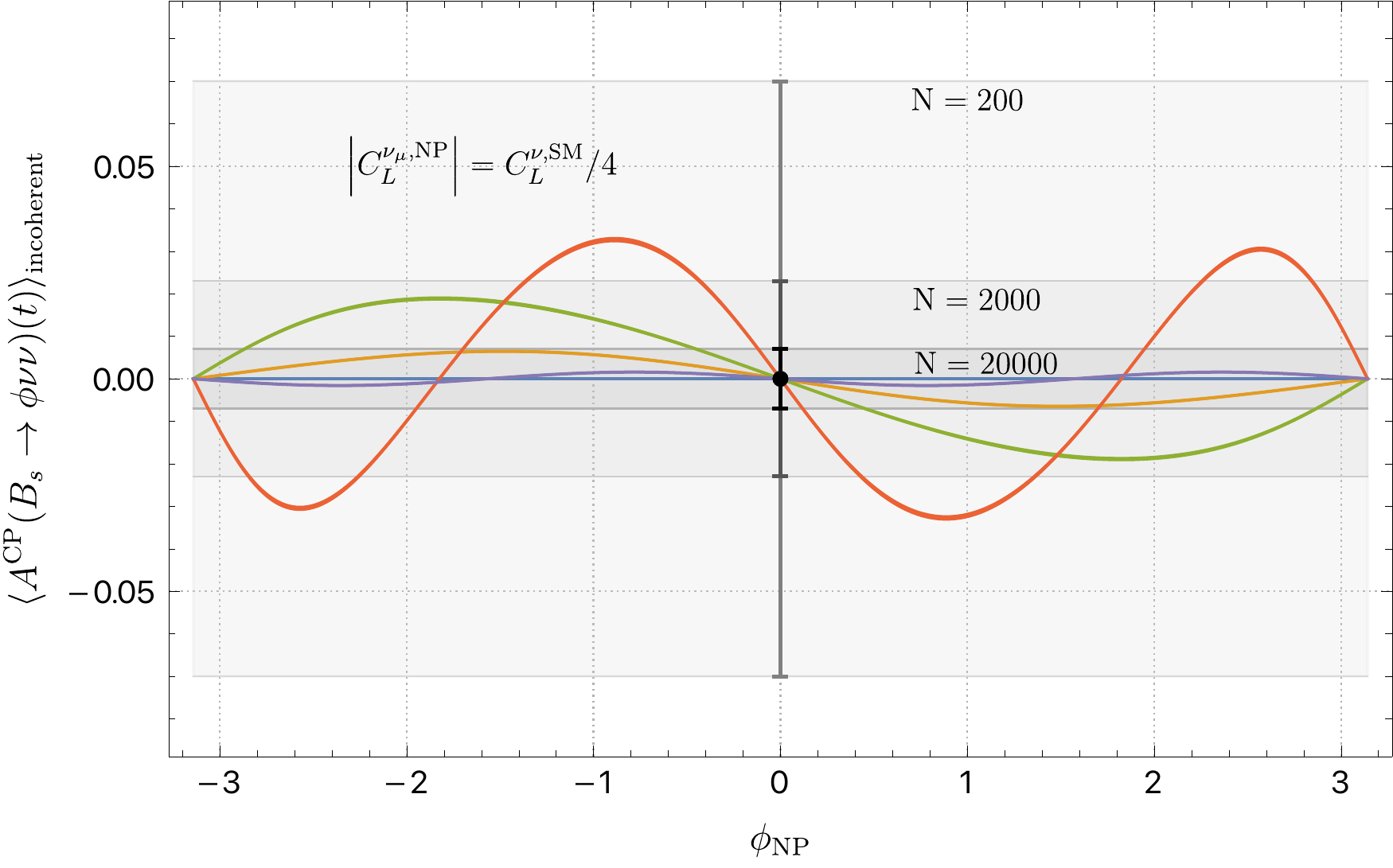}
\includegraphics[width=\columnwidth]{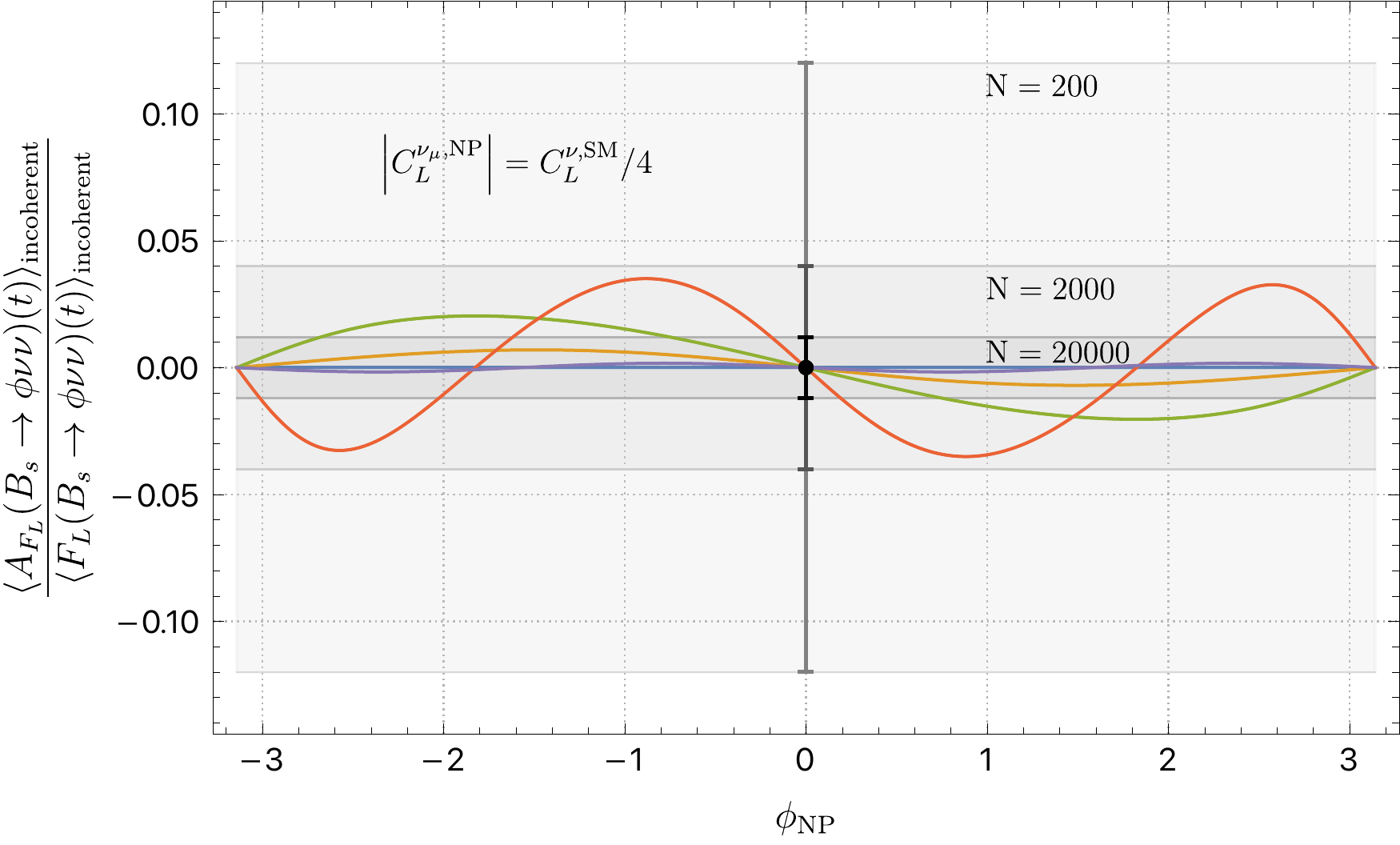}
\caption{Same as \cref{fig:PlotsKS} for the $\langle A_{F_L}\rangle_{\rm incoherent}/\langle F_L\rangle_{\rm incoherent}$ (bottom) $\langle A^{\cal B}\rangle_{\rm incoherent}$ (top) observables of the $B_s\to \phi\nu\bar\nu$ mode. }\label{fig:Plotsphi}
\end{figure}

The procedure used is as follows. Assuming a fixed number of total reconstructed events, we perform a large set of pseudoexperiments, where each corresponds to generating $N$ Monte Carlo events ($N$= 200, 2000, 20000) taking the time-dependent angular differential decay width as probability distribution function (p.d.f.), assuming SM central values. A binned fit on time (and decay angle when pertinent) is then performed with the same p.d.f.~\footnote{We considered several binning choices to check the absence of a binning-induced bias.} The distribution of each of the observables of interest is then studied to obtain the projected statistical uncertainties given in \cref{tab:projections}. Through this approach we neglect effects of detector acceptance, systematic errors and backgrounds that could contaminate this decay, since these effects can only be properly evaluated by experimentalists with a fine understanding of the detector considered and of the distributions of the background events. The projected uncertainties are also shown in \cref{fig:PlotsKS,fig:PlotsKstar,fig:Plotsphi} where they can be compared with the theoretical uncertainties and the dependence on the phase of NP for an arbitrary size for the NP effect as discused previously. 

Concerning the SM predictions, CKM parameters come from CKMFitter 2021~\cite{Charles:2004jd}, form factors for $B\to K_S$ and $B\to K^*$ come from Ref.~\cite{Gubernari:2018wyi}\footnote{Alternative lattice determinations could be used for $B\to K\nu\bar\nu$~\cite{Parrott:2022dnu}, but form factors play a very limited role in the observables discussed in this article.} while for $B_s\to \phi$ form factors come from Ref.~\cite{Bharucha:2015bzk}. 
In  \cref{tab:projections}, we see that the uncertainties are very small for some of the ratios, due to the cancellation of form factor dependencies. We notice
also that some of the observables vanish exactly in the SM, in particular for $B_s^0\to \phi\nu\bar\nu$ where the weak phases in mixing and decay cancel. The phase of $V_{cs}$ is neglected in our analysis (${\rm Im}[V_{cs}]\sim \lambda^6\sim 10^-5$) leading to the vanishing uncertainties of $s_{1c}/J_{1c}$ and $h_{1c}/J_{1c}$. Finally, the direct CP asymmetries $({J_{1s}-\bar{J}_{1s}})/({\Gamma + \bar\Gamma })$ vanish since we assume that there are no NP mechanisms for generating a strong phase.

In Figs.~\ref{fig:PlotsKS}, \ref{fig:PlotsKstar} and \ref{fig:Plotsphi}, we show the variation of some of these observables with $\phi_{\rm NP}$, the global  weak phase involved in all three Wilson coefficients $C^{\nu_\alpha, {\rm NP}}_L$, under the NP scenario $\abs{C^{\nu_\mu, {\rm NP}}_L }=C^{\nu, {\rm SM}}_L/4$. The plots show the SM prediction as well as the values for the four different hypotheses for the lepton couplings (allowing one to derive the NP contributions to the $C^{\nu_e, {\rm NP}}_L$ and $C^{\nu_\tau, {\rm NP}}_L$), as well as the projected experimental statistical uncertainty for 200, 2000 and 20000 events. 

We see that the observables vary significantly with the CP-violating phase $\phi_{\rm NP}$, in particular for scenarios leading to large $\tau$ couplings, whereas they reduce back to the SM values when $\phi_{\rm NP}$ vanishes. We show here a selection of observables or ratios of observables which exhibit small theoretical uncertainties and large variations with the NP phase. From these one can conclude that even experimental measurements based on limited numbers of events could discriminate among the scenarios considered. We stress again that we take into account only statistical uncertainties here, and additional systematics related to specific experimental environments should also be eventually considered in a more detailed dedicated experimental study .

Naturally, these observables can be combined with the branching ratios (and $K^*$ polarisations) discussed in Sec.\ref{sec:btosnunu},  the latter providing constraints on the modulus, and the former probing the phase of the NP contribution, respectively. For the branching ratios, one can estimate that they could be probed down to $\mathcal O(10\%)$ for the 200 event benchmark (assuming $S/B \sim 1$), and $3\%$ (theory limit, based on projected uncertainties of the relevant form factor estimates from Lattice QCD~\cite{Aoki:2021kgd, Kou:2018nap}) for the other two benchmarks. Such a combination is however left for future work, once the experimental perspectives for the new proposed CP-violating observables have been investigated in more detail.

\section{Summary and Conclusions} \label{sec:conclusions}

The hints of NP in $b\to s\ell^+\ell^-$ transitions, potentially involving new weak (CP-violating) phases, motivated us to reconsider the probes of CP-violating NP contributions to $b\to s\nu\bar\nu$. The usual observables (branching ratios and $K^*$ polarisations) in $B\to K(^*)\nu\bar\nu$ all probe the same combinations of the Wilson coefficients $C_L^\nu$ and $C_R^\nu$ in the weak effective Hamiltonian, are sensitive to NP weak phases only in the presence of right-handed currents, and require input concerning the relevant hadronic form factors. We have thus proposed alternative probes of CPV NP by considering the interference between neutral-meson mixing and $b\to s\nu \bar \nu$ decay amplitudes, which are both loop level processes in the SM. We have derived the corresponding observables for the angular analysis $B_d\to K_S\nu\bar\nu$, $B_d\to K^{*0}(\to K_S\pi^0)\nu\bar\nu$, $B_s\to\phi\nu\bar\nu$, exploiting the results already obtained for $b\to s\ell^+\ell^-$ decays in Refs.~\cite{Descotes-Genon:2015hea,Descotes-Genon:2020tnz}. 

It turns out that a few additional angular observables can be obtained either from time-dependent analyses at $B$-factories producing coherent $B$-meson pairs or from time-integrated analyses for machines producing $B$-mesons incoherently, such as the LHC or a $Z$-factory. For $B_d$ decays, one must measure CP-asymmetries, whereas $B_s$ decays allow for the consideration of CP-averaged observables. These observables provide additional information on the phase of the Wilson coefficients $C_L^\nu\pm C_R^\nu$, with limited dependence on the form factors, which furthermore cancels out completely for some ratios of these observables. They can therefore be predicted with very high accuracy.

We have then considered prospective scenarios for Belle II, CEPC and FCC-$ee$ to determine the typical statistical precision that could be reached for the measurements of these observables. 
Within an effective theory description, we have focused on NP scenarios contributing to purely left-handed operators with different couplings but the same arbitrary global phase for all three neutrino flavours. The branching ratios and $K^*$ polarisations are not sensitive to this NP phase. On the other hand, we have shown that several time-dependent and time-integrated observables are highly sensitive to it and a limited number of events (from 200 to 20000) may be enough to exclude some of the NP scenarios considered. We found that both time-dependent analyses at $B$-factories and time-integrated measurements at $Z$-factories could start probing NP scenarios with CP-violating contributions.  Our exploratory study should be improved in the future by including several experimental effects (detector acceptance, backgrounds, systematics), but these first results are rather promising.

To conclude, the patterns of deviations from the SM observed in $b\to s\ell^+ \ell^-$ transitions remain intriguing and require further investigations. The companion $b\to s\nu\bar \nu$ modes should be detected soon at Belle II, and they provide an interesting venue to probe complementary NP contributions in a clean way. We hope that the observables discussed in this article could thus help shed light on the short-distance dynamics of rare $b\to s$ transitions and in particular on its lesser studied CP-violating aspects.

\section*{Acknowledgements}

SF and JFK acknowledge the financial support from the Slovenian Research Agency (research core funding No. P1-0035 and J1-3013).  This project has received support from the European Union’s Horizon 2020 research and innovation programme under the Marie Skodowska-Curie grant agreement No 860881-HIDDeN. This study has been carried out within the INFN project (Iniziativa Specifica) QFT-HEP.

\bibliographystyle{elsarticle-num}

\bibliography{main}

\end{document}